\documentclass[11pt,twoside]{article}
\usepackage{asp2014}

\aspSuppressVolSlug
\resetcounters
\usepackage{natbib}

\begin{document}

\title{Detection of the linearly polarised spectrum of the red supergiant star ${\rm \alpha~Ori}$\\\small\textit{Conference Proceeding of the 8$^{th}$ Solar Polarization Workshop}}
\author{Tessore~B,$^1$ L\'opez Ariste~A,$^2$ Mathias~P,$^2$ Lèbre~A,$^1$ Morin~J$^1$ and Josselin~E,$^1$}
\affil{$^1$LUPM, Universit\'e de Montpellier, CNRS, Place Eug\`ene Bataillon, 34095, France; \email{benjamin.tessore@umontpellier.fr}}
\affil{$^2$Universit\'e de Toulouse, UPS-OMP, Institut de Recherche en Astrophysique et Planétologie, Toulouse, France}

\paperauthor{Tessore~B}{benjamin.tessore@umontpellier.fr}{}{LUPM, Universit\'e de Montpellier, CNRS}{Physique}{Montpellier}{}{34095}{France}
\paperauthor{L\'opez Ariste~A}{arturo.lopezariste@irap.omp.eu}{}{Universit\'e de Toulouse, UPS-OMP, Institut de Recherche en Astrophysique et Planétologie, Toulouse, France}{Physique}{Toulouse}{}{31400}{France}
\paperauthor{Mathias~P}{Philippe.Mathias@irap.omp.eu}{}{Universit\'e de Toulouse, UPS-OMP, Institut de Recherche en Astrophysique et Planétologie, Toulouse, France}{Physique}{Toulouse}{}{31400}{France}
\paperauthor{Josselin~E}{eric.josselin@umontpellier.fr}{}{Universit\'e de Toulouse, UPS-OMP, Institut de Recherche en Astrophysique et Planétologie, Toulouse, France}{Physique}{Toulouse}{}{31400}{France}
\paperauthor{Lèbre~A}{agnes.lebre@umontpellier.fr}{}{LUPM, Universit\'e de Montpellier, CNRS}{Physique}{Montpellier}{}{34095}{France}
\paperauthor{Morin~J}{julien.morin@umontpellier.fr}{}{LUPM, Universit\'e de Montpellier, CNRS}{Physique}{Montpellier}{}{34095}{France}

\begin{abstract}
In the solar limb, linear polarisation is due to anisotropy of the radiation field induced by limb darkening. It is maximal when it is seen parallel to the limb and it vanishes when it is integrated over the spherically-symmetric solar disk.
Therefore for distant stars, that present spherical symmetry, linear polarisation signatures are very difficult to observe.
 However strong linear polarisation features have been reported in the prototypical red supergiant star ${\rm \alpha~Ori}$ (Betelgeuse). With an analytical model we propose to explain them.
 
\end{abstract}

\section{Introduction}
Red supergiant stars (M type supergiants, hereafter RSG) are cool (${\rm T_{eff}}$ between 3,000 K and 4,000K) and massive (M ${\rm > 10~M_{\odot}}$) stars. 
They have a very extended atmosphere (from ${\rm 100R_{\odot}~to~100000R_{\odot}}$ for the biggest ones) rich in molecules and dust grains. 
They undergo an important mass loss (${\rm 10^{-6}~-~10^{-4}~M_{\odot}/yr}$) and are therefore among the main recycling agents of the interstellar medium. 
While the mass loss is a key ingredient in stellar evolution models \citep{2013EAS....60...31E} it is still poorly understood and the mechanisms triggering it are not well constrained.
\cite{2007A&A...469..671J} proposed that a vigorous convection, taking place at the base of the extended atmosphere (hereafter, the photosphere), could explain how a mass loss event starts.
It is therefore important to study the photosphere of RSG and to characterise its dynamics. Spectroscopic and interferometric observations in the visible and near infra-red have been widely used to study the photosphere and are in good agreement with hydrodynamic simulations (MHD).
Indeed, MHD simulations of \cite{2002AN....323..213F} and \cite{2011A&A...535A..22C} predict that giant convective cells lie above the photosphere, with sizes of about 10\% of the stellar radii. 
Moreover, \cite{2010A&A...516L...2A}, using spectropolarimetry, have detected a weak magnetic field at the surface of ${\rm \alpha~Ori}$. Because of the very long rotation period of the stars (of about 15 years) a small-scale dynamo generating a global magnetic field is supposed.
This detection comforts the idea that turbulent motions in the giant convective cells may generate a global magnetic field \citep[see for instance][]{2004A&A...423.1101D}.
From recent observations of RSG stars, \cite{2016A&A...591A.119A} and Tessore et al. (in preparation) have shown that spectropolarimetry, taking advantage of the great quantity of informations tangled in linear polarisation, could be a promising tool to study the surface dynamics of these evolved stars.

\section{Observations and data analysis}
${\rm \alpha~Ori}$ is the brightest and closest RSG. It has naturally deserved many observations (with spectroscopy and interferometry) in all wavelength ranges so that it is possible to have an extensive overview of the star itself and to link one part of its extended atmosphere with another. ${\rm \alpha~Ori}$ belongs to an observing program we have initiated in 2015 at Télescope Bernard-Lyot (TBL, Pic du Midi France) with the Narval instrument, the twin of the spectropolarimeter ESPaDOnS \citep{2006ASPC..358..362D}.
This large program is dedicated to the spectropolarimetric study of cool evolved stars using both circular and linear polarisation spectra.
Each polarimetric sequence (one spectrum) collected by Narval covers a spectral range from 375 nm to 1,050 nm in a single exposure with a resolving power of about 65,000. 
The optimal extraction of spectra is performed with Libre-ESpRIT \citep{1997MNRAS.291..658D}, an automatic and dedicated reduction package installed at TBL and it includes wavelength calibration, continuum normalisation and correction to the heliocentric rest frame.
A Null spectrum is also computed for each observations. This Null diagnosis has no physical meaning but in absence of spurious polarisation it is featureless.
In all linear polarised observations of ${\rm \alpha~Ori}$ the Null signal is always flat at the noise level, this indicates the good quality of the polarimetric observations as well as the stellar origin of the detected polarised features.
Since March 2015 the star has been observed once per month and eight Stokes Q and Stokes U spectra are collected for each observational date.

\begin{figure}[h!]
\begin{center}
 \includegraphics[scale=0.55]{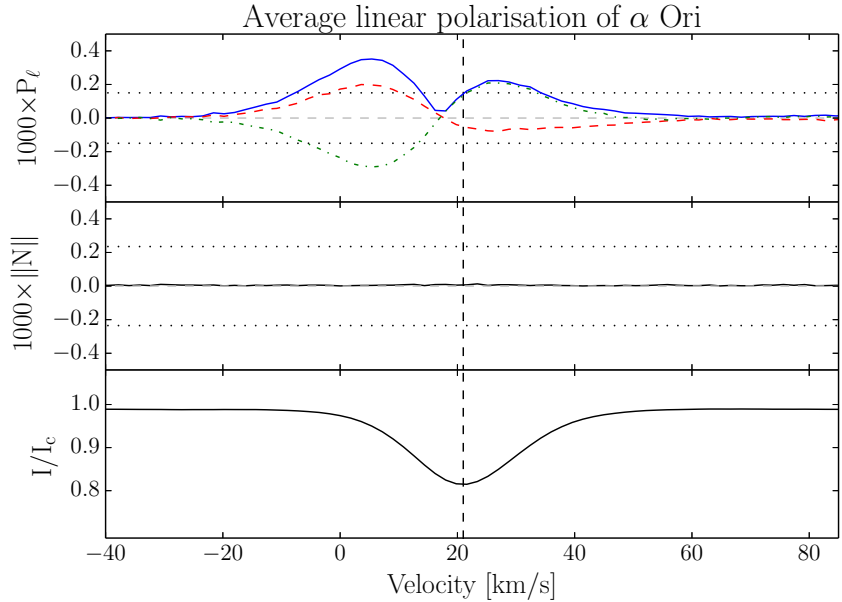}
 \caption{Average LSD profile of ${\rm \alpha~Ori}$ for the observations of  2015 - 2016. The bottom panel shows the mean line profile and the middle panel shows the mean Null signal. The top panel shows the 
 mean linear polarisation profile with Stokes Q (red dashed line), Stokes U (green dot dashed line) and the total amount of linear polarisation (blue solid line). The vertical line indicates the velocity of the star in the heliocentric rest frame. The error bars are The error bars are represented by the dotted horizontal lines in the top and middle panels and they are scaled by a factor 100.}
 \label{lsd_prof}
\end{center}
\end{figure}

Furthermore we have analysed each spectrum with the Least-Square Deconvolution (LSD) method \citep{1997MNRAS.291..658D}.
The LSD method extracts from thousands of spectral lines a mean information called LSD profile. \\Each LSD profile is a set of a mean line profile, mean polarimetric profile and the mean Null signal (${\rm I/Ic,~Q/Ic,~U/Ic,~N~and~P_{\ell}=\sqrt(Q^2 + U^2)}$). Figure \ref{lsd_prof} shows a typical LSD profile of ${\rm \alpha~Ori}$. 

\section{Linearly polarised spectrum of ${\rm \alpha~Ori}$}
\begin{figure}[h!]

 \includegraphics[scale=0.3]{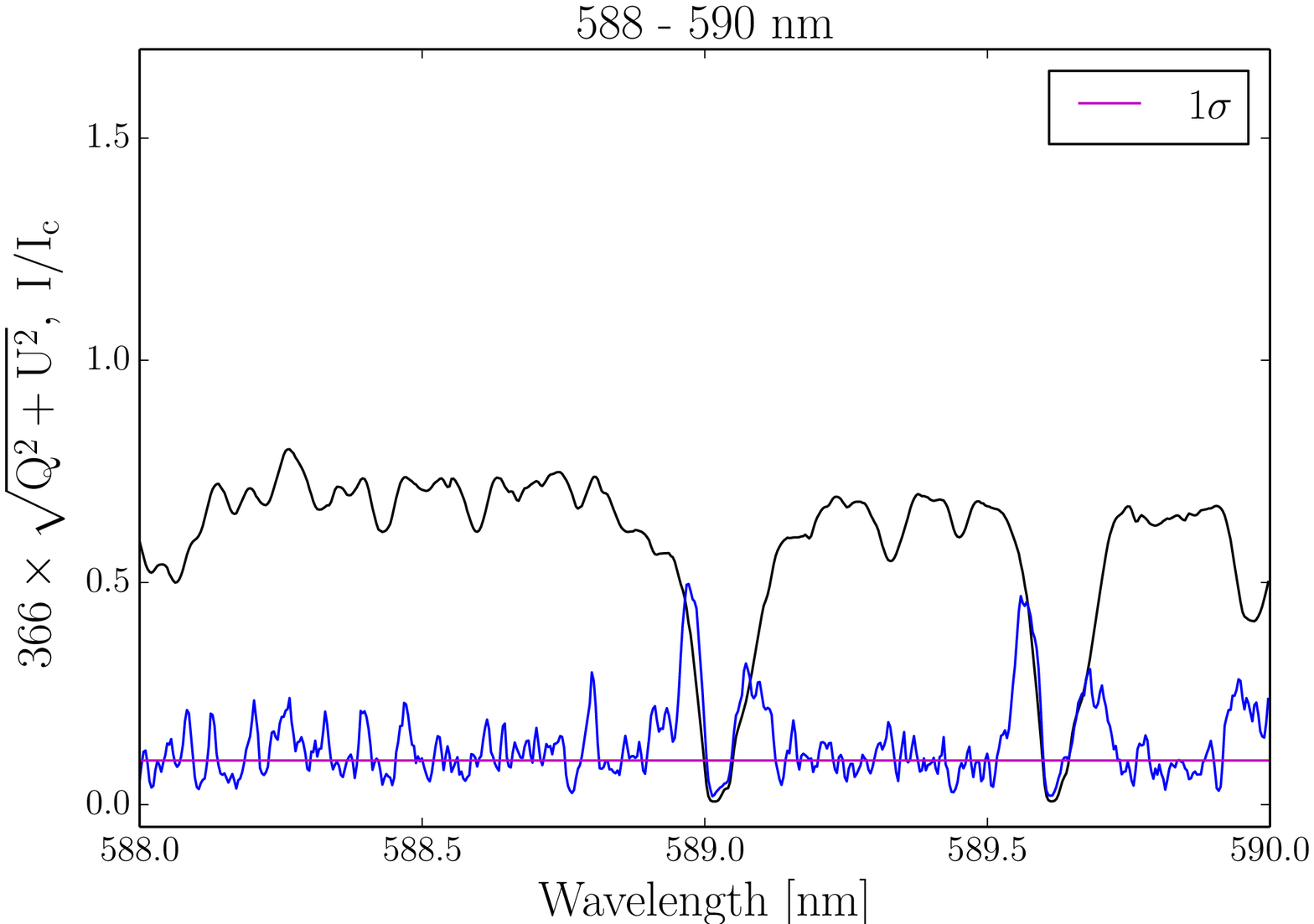}
  \includegraphics[scale=0.3]{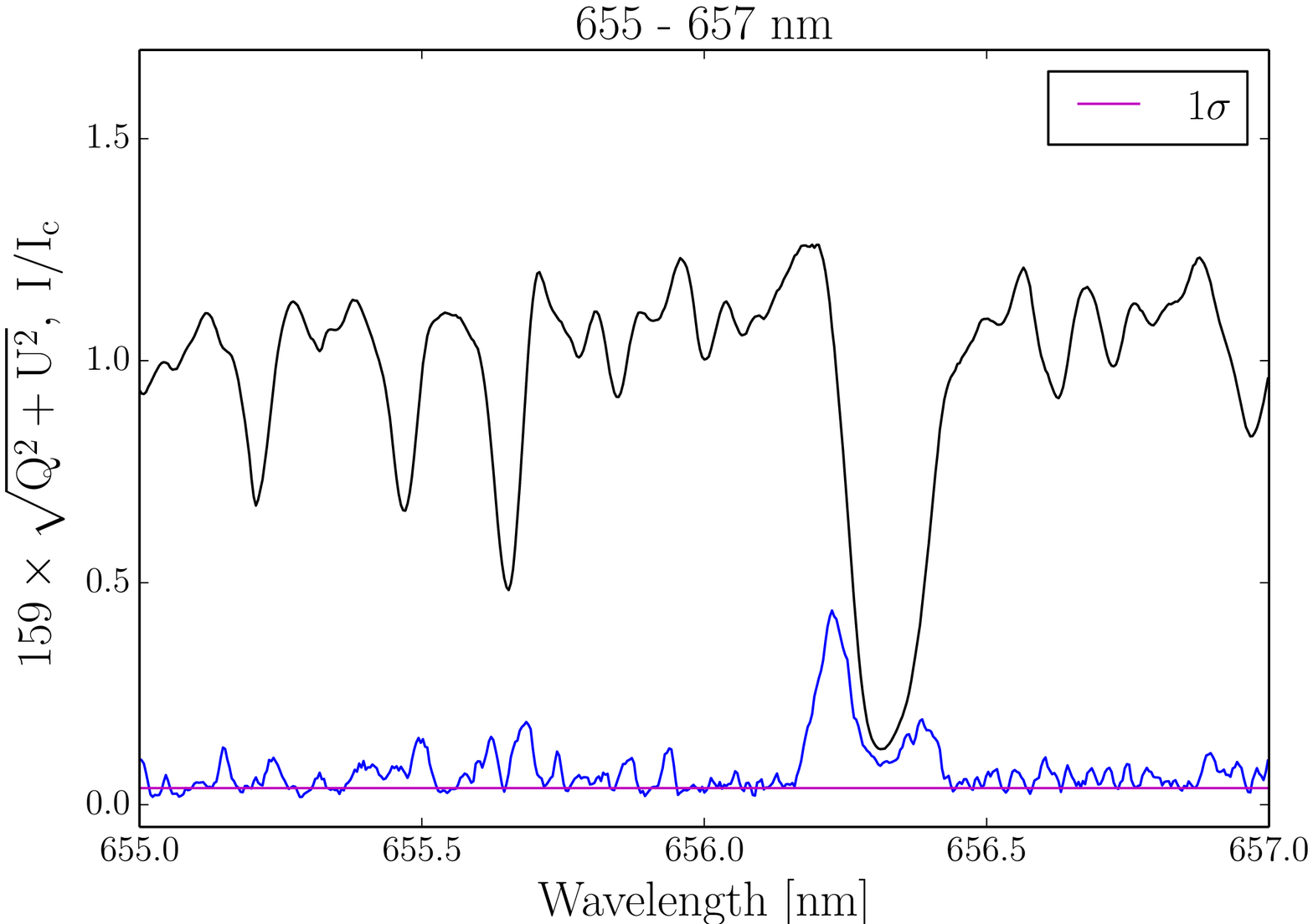}

 \caption{Linearly polarised spectrum of ${\rm \alpha~Ori}$ around the sodium D lines (left panel) and ${\rm H_{\alpha}}$ line (right panel).
        The black solid lines (upper part of each panel) are the unpolarised flux (Stokes I). The blue solid lines (lower part of each panel) are the total amount of linear polarisation. The 1${\rm \sigma}$ level is shown by the horizontal lines.}
 \label{linearsp}
 
\end{figure}

\begin{figure}[h!]
  \begin{center}

 \includegraphics[scale=0.55]{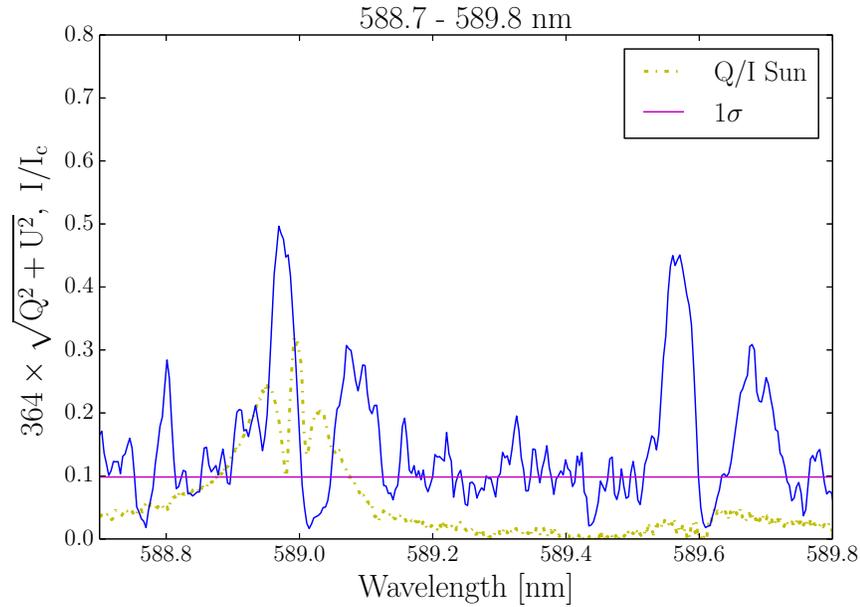}
 
 \caption{The Sodium D lines in the linearly polarised spectrum of ${\rm \alpha~Ori}$ (solid line) as compared to the solar case (dot-dashed line). The horizontal line is the 1${\rm \sigma}$ level of the observation of ${\rm \alpha~Ori}$. No scaling factor has been applied to the second solar spectrum. The solar data are taken from \url{www.irsol.ch/data-archive/} \citep{2000sss..book.....G}}
 \label{Dlines_alpOri}
 \end{center}
\end{figure}

The complex linearly polarised spectrum of the Sun, known as the "Second Solar Spectrum" \citep{1997A&A...321..927S} has two main contributions. A few lines show polarisation such as the Strontium line at 460.7 nm and the Sodium D lines around 588.9 nm which can be explained by complex quantum interferences while the majority of the lines depolarise the polarised continuum spectrum. At the depth where the continuum get polarised, polarised photons (and unpolarised ones) are emitted. These photons are absorbed by the surrounding stellar gas and re-emitted, unpolarised, during the line forming process. Therefore at the frequencies of the lines the continuum polarisation presents a depression having the shape of the lines. The continuum spectrum of the Sun gets polarised by Rayleigh scattering at neutral Hydrogen and Thomson scattering at free electrons. This polarisation is modified by the geometry of the radiation field that defines how the atoms are illuminated. In the Sun, the centre-to-limb variation (CLV) of the intensity introduces anisotropy in the radiation field that is stronger near the limb. \\Red supergiant stars also present polarisation of their continuum. \cite{1986ApJ...307..261D} has measured the continuum polarisation of ${\rm \alpha~Ori}$ at 0.5\% in the blue. Because RSG are cooler than the sun, the main contributors to the continuum polarisation are Hydrogen atoms and molecules. 
\\For symmetry reasons, looking at the Sun as a star, the Second Solar Spectrum would not be observable. It is therefore very difficult to observe such a polarised spectrum for distant stars. However the strong brightness contrasts introduced by the giant convective cells lying at the surface of RSG are likely sufficient to break the spherical symmetry of the stellar disk, so that after integrating over the stellar disk a little fraction of the continuum polarisation may remain.
\cite{2016A&A...591A.119A} have detected a complex linearly polarised spectrum of ${\rm \alpha~Ori}$. Figure \ref{linearsp} shows the linearly polarised spectrum of ${\rm \alpha~Ori}$ around two different regions.
Because the surface magnetic field  of ${\rm \alpha~Ori}$ is at the Gauss level, \cite{2016A&A...591A.119A} have disregarded a Zeeman origin of the linear polarisation. Furthermore, by looking at the Sodium D lines in ${\rm \alpha~Ori}$ and comparing it with the same lines in the Sun (see Fig. \ref{Dlines_alpOri}) \cite{2016A&A...591A.119A} have shown that depolarisation of continuum dominates the linearly polarised spectrum of ${\rm \alpha~Ori}$. This explains why there are many features in the polarised spectrum of ${\rm \alpha~Ori}$ facing the intensity lines and showing a similar shape. Indeed, \cite{2003A&A...398..763F} explained that the depolarising lines have statistically the same shape (that is why the LSD methods is well suitable).\\At this time we know that the continuum of ${\rm \alpha~Ori}$ (and RSG) is polarised by Rayleigh scattering and that we mainly observed depolarisation by the photospheric lines.

\section{Spectropolarimetry as a diagnostic tool to study the surface dynamics of red supergiant stars}

\begin{figure}[h!]
  \begin{center}
  \includegraphics[scale=0.55]{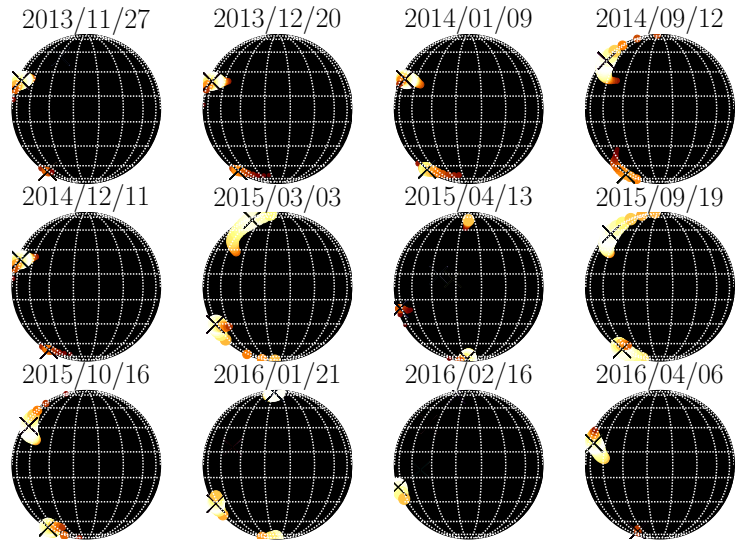}
  \caption{Reconstructed positions of bright spots at the surface of ${\rm \alpha~Ori}$ along the observation period. The crosses indicate the maximum value of brightness for each spot.}
  \label{map}
 \end{center}
 \end{figure}
 
As explained in the previous sections the photosphere of RSG is very complex. The brightness contrasts introduced by the giant convective cells are responsible for the small changes in the luminosity of the star. \cite{2009A&A...508..923H} from interferometric observations have reconstructed two bright spots at the surface of ${\rm \alpha~Ori}$. More recently, \cite{2015EAS....71..243M} using their interferometric model have inferred two Gaussian spots at the surface of ${\rm \alpha~Ori}$.
From the point of view of the observations there is a clear link between the surface convective patterns of the star and the measured linear polarisation. Therefore, \cite{2016A&A...591A.119A} have developed an analytical model that link the observed LSD profiles of ${\rm \alpha~Ori}$ to bright spots at the surface of the star. The main idea of this model is that each maximum in the LSD profile corresponds to one bright spot at the surface of the star. This spot is in expansion if it is shifted to the blue in the LSD profile or sinking if it is shifted to the red. Therefore the amplitudes of Stokes Q and Stokes U and their positions give us the polar coordinates of the spots, ${\rm \theta~and~\mu}$. However, the exact dependence of the polarisation with the intrinsic brightness of each spot is not known and only the ratio of the polarisation between each spot with respect to the brightest one, corrected by the angular dependence of Rayleigh scattering, is given. For each observation, for each LSD profile the positions of several spots are inferred on the disk of ${\rm \alpha~Ori}$. Figure \ref{map} shows the possible positions of the bright spots inferred by the model for ${\rm \alpha~Ori}$ using all the spectropolarimetric observations collected on this star since November 2013.
\\Remarkably the spectropolarimetric model of spots described by of \cite{2016A&A...591A.119A} is in good agreement with the interferometric model of \cite{2015EAS....71..243M}. Although this analytical spectropolarimetric model does not reproduce a continuous distribution of brightness (more realistic) and does not provide absolute values of polarisation it is still a good tool to study the time variation of the photosphere and it has proven is ability to complement interferometric observations.

Therefore with these preliminary results we have shown how spectropolarimetry could be a great tool in characterising the photosphere of red supergiant stars. New ongoing observations of ${\rm \alpha~Ori}$ as well as other RSG such as CE Tau and ${\rm \mu~Cep}$ will help in improving the model and will open a new way to study convection and mass loss in massive evolved stars.
Moreover, new joint Narval observations with interferometric observations are planed.

 \bibliographystyle{asp2014}
 \bibliography{tessore_1.bib}  

\begin{thebibliography}{}
\expandafter\ifx\csname natexlab\endcsname\relax\def\natexlab#1{#1}\fi
\expandafter\ifx\csname url\endcsname\relax
  \def\url#1{\texttt{#1}}\fi
\expandafter\ifx\csname urlprefix\endcsname\relax\def\urlprefix{URL }\fi
\providecommand{\eprint}[2][]{\url{#2}}

\bibitem[{{Auri{\`e}re} et~al.(2010){Auri{\`e}re}, {Donati},
  {Konstantinova-Antova}, {Perrin}, {Petit}, \&
  {Roudier}}]{2010A&A...516L...2A}
{Auri{\`e}re}, M., {Donati}, J.-F., {Konstantinova-Antova}, R., {Perrin}, G.,
  {Petit}, P., \& {Roudier}, T. 2010, \aap, 516, L2. \eprint{1005.4845}

\bibitem[{{Auri{\`e}re} et~al.(2016){Auri{\`e}re}, {L{\'o}pez Ariste},
  {Mathias}, {L{\`e}bre}, {Josselin}, {Montarg{\`e}s}, {Petit}, {Chiavassa},
  {Paletou}, {Fabas}, {Konstantinova-Antova}, {Donati}, {Grunhut}, {Wade},
  {Herpin}, {Kervella}, {Perrin}, \& {Tessore}}]{2016A&A...591A.119A}
{Auri{\`e}re}, M., {L{\'o}pez Ariste}, A., {Mathias}, P., {L{\`e}bre}, A.,
  {Josselin}, E., {Montarg{\`e}s}, M., {Petit}, P., {Chiavassa}, A., {Paletou},
  F., {Fabas}, N., {Konstantinova-Antova}, R., {Donati}, J.-F., {Grunhut},
  J.~H., {Wade}, G.~A., {Herpin}, F., {Kervella}, P., {Perrin}, G., \&
  {Tessore}, B. 2016, \aap, 591, A119. \eprint{1605.04702}

\bibitem[{{Chiavassa} et~al.(2011){Chiavassa}, {Freytag}, {Masseron}, \&
  {Plez}}]{2011A&A...535A..22C}
{Chiavassa}, A., {Freytag}, B., {Masseron}, T., \& {Plez}, B. 2011, \aap, 535,
  A22. \eprint{1109.3619}

\bibitem[{{Doherty}(1986)}]{1986ApJ...307..261D}
{Doherty}, L.~R. 1986, \apj, 307, 261

\bibitem[{{Donati} et~al.(2006){Donati}, {Catala}, {Landstreet}, \&
  {Petit}}]{2006ASPC..358..362D}
{Donati}, J.-F., {Catala}, C., {Landstreet}, J.~D., \& {Petit}, P. 2006, in
  Astronomical Society of the Pacific Conference Series, edited by R.~{Casini},
  \& B.~W. {Lites}, vol. 358 of Astronomical Society of the Pacific Conference
  Series, 362

\bibitem[{{Donati} et~al.(1997){Donati}, {Semel}, {Carter}, {Rees}, \& {Collier
  Cameron}}]{1997MNRAS.291..658D}
{Donati}, J.-F., {Semel}, M., {Carter}, B.~D., {Rees}, D.~E., \& {Collier
  Cameron}, A. 1997, \mnras, 291, 658

\bibitem[{{Dorch}(2004)}]{2004A&A...423.1101D}
{Dorch}, S.~B.~F. 2004, \aap, 423, 1101. \eprint{astro-ph/0403321}

\bibitem[{{Ekstr{\"o}m} et~al.(2013){Ekstr{\"o}m}, {Georgy}, {Meynet}, {Groh},
  \& {Granada}}]{2013EAS....60...31E}
{Ekstr{\"o}m}, S., {Georgy}, C., {Meynet}, G., {Groh}, J., \& {Granada}, A.
  2013, in EAS Publications Series, edited by P.~{Kervella}, T.~{Le Bertre}, \&
  G.~{Perrin}, vol.~60 of EAS Publications Series, 31. \eprint{1303.1629}

\bibitem[{{Fluri} \& {Stenflo}(2003)}]{2003A&A...398..763F}
{Fluri}, D.~M., \& {Stenflo}, J.~O. 2003, \aap, 398, 763

\bibitem[{{Freytag} et~al.(2002){Freytag}, {Steffen}, \&
  {Dorch}}]{2002AN....323..213F}
{Freytag}, B., {Steffen}, M., \& {Dorch}, B. 2002, Astronomische Nachrichten,
  323, 213

\bibitem[{{Gandorfer}(2000)}]{2000sss..book.....G}
{Gandorfer}, A. 2000, {The Second Solar Spectrum: A high spectral resolution
  polarimetric survey of scattering polarization at the solar limb in graphical
  representation. Volume I: 4625 {\AA} to 6995 {\AA}}

\bibitem[{{Haubois} et~al.(2009){Haubois}, {Perrin}, {Lacour}, {Verhoelst},
  {Meimon}, {Mugnier}, {Thi{\'e}baut}, {Berger}, {Ridgway}, {Monnier},
  {Millan-Gabet}, \& {Traub}}]{2009A&A...508..923H}
{Haubois}, X., {Perrin}, G., {Lacour}, S., {Verhoelst}, T., {Meimon}, S.,
  {Mugnier}, L., {Thi{\'e}baut}, E., {Berger}, J.~P., {Ridgway}, S.~T.,
  {Monnier}, J.~D., {Millan-Gabet}, R., \& {Traub}, W. 2009, \aap, 508, 923.
  \eprint{0910.4167}

\bibitem[{{Josselin} \& {Plez}(2007)}]{2007A&A...469..671J}
{Josselin}, E., \& {Plez}, B. 2007, \aap, 469, 671. \eprint{0705.0266}

\bibitem[{{Montarg{\`e}s} et~al.(2015){Montarg{\`e}s}, {Kervella}, {Perrin},
  {Chiavassa}, \& {Auri{\`e}re}}]{2015EAS....71..243M}
{Montarg{\`e}s}, M., {Kervella}, P., {Perrin}, G., {Chiavassa}, A., \&
  {Auri{\`e}re}, M. 2015, in EAS Publications Series, vol.~71 of EAS
  Publications Series, 243. \eprint{1512.03314}

\bibitem[{{Stenflo} \& {Keller}(1997)}]{1997A&A...321..927S}
{Stenflo}, J.~O., \& {Keller}, C.~U. 1997, \aap, 321, 927

\end{thebibliography}

\end{document}